\documentclass[prx,twocolumn,superscriptaddress,floatfix,nopacs,nofootinbib]{revtex4-1}
\usepackage{graphicx,amsfonts,amssymb,amsmath,hyperref,enumerate}

\newif\ifhyper
\hypertrue
\ifhyper
\hypersetup{
   citecolor = {green},
   colorlinks = {true}, 
   urlcolor = {blue} 
}

\newcommand{\beq}{\begin{equation}}
\newcommand{\eeq}{\end{equation}}
\newcommand{\beqa}{\begin{eqnarray}}
\newcommand{\eeqa}{\end{eqnarray}}
\newcommand{\ket} [1] {\vert #1 \rangle}

\def\ket#1{\vert#1\rangle}

\def\Longarrow{\protect\@lra}
\def\@lra{\relbar\joinrel\relbar\joinrel\relbar\joinrel%
          \relbar\joinrel\rightarrow}

\begin{document} 

\title{Quantum artificial vision for defect detection in manufacturing}

\author{Daniel Guijo}
\affiliation{Multiverse Computing, Paseo de Miram\'on 170, E-20014 San Sebasti\'an, Spain}

\author{Victor Onofre}
\affiliation{Multiverse Computing, Paseo de Miram\'on 170, E-20014 San Sebasti\'an, Spain}

\author{Gianni Del Bimbo}
\affiliation{Multiverse Computing, Paseo de Miram\'on 170, E-20014 San Sebasti\'an, Spain}

\author{Samuel Mugel}
\affiliation{Multiverse Computing, Centre for Social Innovation, 192 Spadina Ave, Suite 412, Toronto M5T 2C2, Canada}

\author{{Daniel Estepa}}
\affiliation{Ikerlan Technology Research Centre, Basque Research and Technology Alliance (BRTA), Paseo J.M. Arizmediarrieta 2, E-20500, Arrasate-Mondrag\'on, Spain}
\author{{Xabier De Carlos}}
\affiliation{Ikerlan Technology Research Centre, Basque Research and Technology Alliance (BRTA), Paseo J.M. Arizmediarrieta 2, E-20500, Arrasate-Mondrag\'on, Spain}
\author{{Ana Adell}}
\affiliation{Ikerlan Technology Research Centre, Basque Research and Technology Alliance (BRTA), Paseo J.M. Arizmediarrieta 2, E-20500, Arrasate-Mondrag\'on, Spain}
\author{{Aizea Lojo}}
\affiliation{Ikerlan Technology Research Centre, Basque Research and Technology Alliance (BRTA), Paseo J.M. Arizmediarrieta 2, E-20500, Arrasate-Mondrag\'on, Spain}
\author{{Josu Bilbao}}
\affiliation{Ikerlan Technology Research Centre, Basque Research and Technology Alliance (BRTA), Paseo J.M. Arizmediarrieta 2, E-20500, Arrasate-Mondrag\'on, Spain}

\author{Rom\'an Or\'us}
\affiliation{Multiverse Computing, Paseo de Miram\'on 170, E-20014 San Sebasti\'an, Spain}
\affiliation{Donostia International Physics Center, Paseo Manuel de Lardizabal 4, E-20018 San Sebasti\'an, Spain}
\affiliation{Ikerbasque Foundation for Science, Maria Diaz de Haro 3, E-48013 Bilbao, Spain}

\begin{abstract}
In this paper we consider several algorithms for quantum computer vision using Noisy Intermediate-Scale Quantum (NISQ) devices, and benchmark them for a real problem against their classical counterparts. Specifically, we consider two approaches: a quantum Support Vector Machine (QSVM) on a universal gate-based quantum computer, and QBoost on a quantum annealer. The quantum vision systems are benchmarked for an unbalanced dataset of images where the aim is to detect defects in manufactured car pieces. We see that the quantum algorithms outperform their classical counterparts in several ways, with QBoost allowing for larger problems to be analyzed with present-day quantum annealers. Data preprocessing, including dimensionality reduction and contrast enhancement, is also discussed, as well as hyperparameter tuning in QBoost. To the best of our knowledge, this is the first implementation of quantum computer vision systems for a problem of industrial relevance in a manufacturing production line.  
\end{abstract}

\maketitle

\section{Introduction}
\label{sec1}
We are living very interesting times for quantum computing. Some quantum hardware proposals are getting to a point where one can actually start using them for tasks that are classically very hard \cite{quantumsupremacy}. One can, in fact, start thinking about industrial applications of these machines \cite{industrial}. The plans to increase the capabilities of quantum hardware in the forthcoming years are promising, including proposals for scalability and error-correction. All in all, quantum computing is starting to become relevant not just for science, but also for industry: the technology, even if still in its childhood, is starting to hit the market. 

An important field of application of quantum technologies is \emph{manufacturing}. In this vertical, one expects many applications of quantum computing, both for optimization and machine learning, including predictive maintenance, artificial vision, optimization of factory logistics, and more. Among these, the problem of detecting defects in manufactured pieces is of high practical value. This is a typical problem in artificial vision related to the quality assessment of a production line, and companies cope nowadays with several approaches to this problem. The most modern approaches use advanced AI-based artificial vision methods to identify such defects. Classical approaches for computer vision are dominated by the use of deep Convolutional Neural Networks (CNN) which, due to the huge amount of parameters they contain, act as black boxes and result in high inference times caused by the number of operations needed for just a single forward pass.

In this paper we explore the capabilities of noisy intermediate-scale quantum (NISQ) hardware for the task of binary image classification using quantum machine learning (QML) techniques, in the context of defect detection. What we propose here is \emph{to use algorithms for artificial vision on quantum computers}. In particular, we propose two approaches: a quantum Support Vector Machine (QSVM) for a universal gate-based quantum computer, as well as QBoost on a quantum annealer. We benchmark the proposed quantum vision systems against their classical counterparts for an unbalanced dataset of images from the GDXray+ public dataset \cite{gdxray} consisting of fractures in manufactured car pieces. In the end, our conclusion is that quantum vision systems outperform their classical counterparts, being able to detect defects with higher precision. As we shall see, this is particularly relevant for the QBoost algorithm when implemented on current quantum annealers, being able to deal with problems of industrially-relevant size without significant data preprocessing. 

This paper is organized as follows: in Sec.\ref{sec2} we explain the problem setting, including the dataset, image preprocessing, and relevant metrics. In Sec.\ref{sec3} the methods are introduced, namely QSVM, QBoost, and their classical counterparts. Sec.\ref{sec4} deals with the results, including information on dimensionality reduction, contrast enhancement, and hyperparameter tuning in QBoost. Finally, Sec.\ref{sec5} wraps up with the conclusions.  

\section{Problem setting}
\label{sec2}

Let us start by explaining the considered image dataset, as well as the specific task to be implemented, including data preprocessing and relevant metrics.

\subsection{Dataset}

\begin{figure}
\centering
\includegraphics[width=\linewidth]{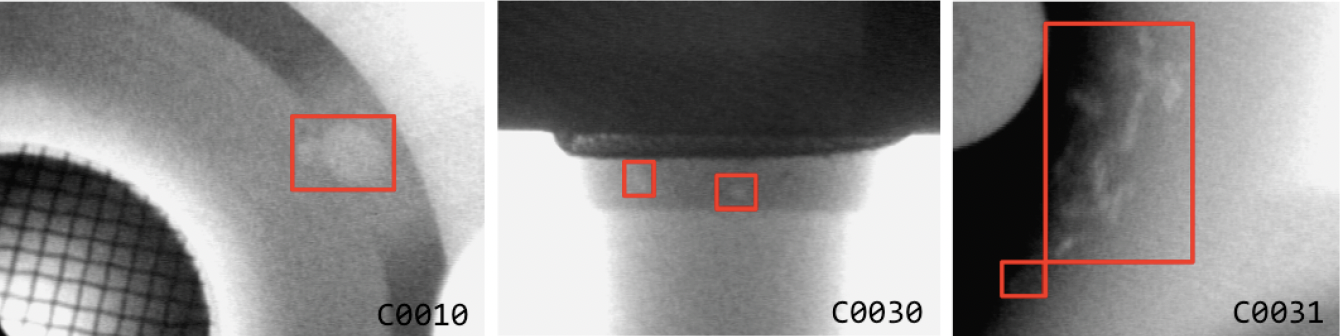}
\caption{[Color online] Examples of images of the ``Castings" group, with the corresponding bounding boxes surrounding the defects. Picture taken from Ref.\cite{gdxray}, reproduced with permission.}
\label{defectfig}
\end{figure}

The GDXray+ dataset \cite{gdxray} is a public X-ray dataset for computer vision testing and evaluation. The dataset consist on five different groups of images, of which only the first one has been used. This group, called ``Castings", contains 67 series with a total of 2727 images of automotive parts with and without casting defects, and annotations of the bounding boxes of these defects. Examples of the images are shown in Fig. \ref{defectfig}.

We labeled the images as ``with defect" or ``without defect", which a priori is a simpler scenario than adding bounding boxes. The labeling was done using a custom-made program written in Python, and based on the existence of bounding box annotations for each image. The goal is then to train a supervised classification system that is able to \emph{see} whether an image has a defect or not\footnote{This basic task can, in fact, be subsequently refined in order to actually identify the position of the defect in the image.}. Due to the varying size of the images of different series (e.g. 439.3 kpixels for series C0001 and 44.5 kpixels for series C0002), and the existence of low contrast defects in some of the images, several techniques have been used to preprocess the data before feeding it to the classifiers. This is explained in what follows. 

\subsection{Preprocessing}

As mentioned before, the images vary in size between series, and so they were reshaped to 320x428 and flattened, in addition to standarization and normalization procedures. Some of the images present low contrast defects (again, see Fig. \ref{defectfig}), and therefore one may use different contrast enhancement techniques to highlight the visibility of these features. Due to the low number of qubits on current gate-based quantum hardware, dimensionality reduction is imperative in order to feed any model implemented in these  devices. With this in mind, results applying \emph{Principal Component Analysis} (PCA) are presented in Sec.\ref{sec4}.

PCA is a dimensionality reduction technique which allows to simplify the complexity of spaces with a high number of dimensions while preserving the amount of information. These principal components are obtained by finding the vectors in which direction the variance of the data is maximum, that is, the \emph{eigenvectors}. Once the first principal component has been found, we impose that the next one should be orthogonal to the first one to ensure linear independence between components. This second component would correspond to that with the direction, orthogonal to the first one, with the second highest variance, which would correspond to the eigenvector with the second highest \emph{eigenvalue} (i.e. length). We repeat this process iteratively to obtain the rest of the principal components.

\subsection{Metrics}

The metrics used during this study are three of the most used and common in machine learning and data science, namely \emph{Precision}, \emph{Recall} and \emph{F1-Score}. Precision is defined as the number of true positives over the sum of the number of true and false positives:
$$P=\frac{T_{P}}{T_{P}+F_{P}}.$$
Intuitively, this can be seen as the ability of the model to not classify as positive a sample that is negative. Recall is defined as the number of true positives over the sum of the number of true positives and false negatives:
$$R=\frac{T_{P}}{T_{P}+F_{N}}.$$
This can be understood as the ability of the model to find all the positive samples. Since there tends to be a trade-off between these two metrics, specially in unbalanced datasets, we also use the F1-Score of the classifiers to provide a more consistent metric. This is  defined as the harmonic mean of precision and recall:
$$F1=2\cdot\frac{P\cdot R}{P+R}.$$
This metric cannot be greater than the highest of recall and precision, and it cannot be smaller than the lowest of the two. Besides, it is closer to the lower one, meaning that the F1 score cannot be greater that the arithmetic mean. These two characteristics make that the F1 score enforces the classification algorithm to be good in both recall and precision, unifying them in only one metric.

\section{Methods}
\label{sec3}

To asses the capabilities of current quantum hardware, we explored two different models which are implemented on different devices: a \textit{Quantum Support Vector Machine} (QSVM) run on a simulator of IBM quantum hardware, and the $QBoost$ algorithm run on a \textit{D-Wave} quantum annealer. This was benchmarked against several classical methods, which we also explain below. 

\subsection{Quantum Support Vector Machine on a universal quantum computer}

As proposed in Havlicek et al. \cite{havlicek}, the QSVM is a \textit{quantum-enhanced} kernel method. Instead of performing the classification task on the original space of the data, each datapoint is mapped onto a Hilbert space of $n$ qubits and with dimension $2^{n}$, by a \textit{quantum feature map} $\psi$, where $n$ is also the number of features in each datapoint. In fact, via the feature map, a data point $\vec{x}$ in the feature space is mapped to a quantum state $\ket{\psi(\vec{x})}$ of $n$ qubits. The goal of this transformation is to ease the task of finding the separating hyperplane between the two classes using a classical SVM algorithm, and this problem boils down to calculating the distances between each pair of datapoints in the new feature space. The distance between two vectors $\vec{x}_{1}$ and $\vec{x}_{2}$ in this space is calculated via the inner product:
$$K(\vec{x}_{1}, \vec{x}_{2})=\vert\langle \psi(\vec{x}_{1})\vert \psi(\vec{x}_{2})\rangle\vert^{2}.$$ This set of distances forms the \textit{quantum kernel matrix}, $K$, of the transformation. The higher dimensionality of this Hilbert space, together with the information of correlations between the features provided by quantum entanglement, ease the classification task for the SVM algorithm.

For our problem, we used the \textit{ZZ Feature Map} \cite{yano} provided by IBM's  QSDK, \textit{Qiskit}, which can be implemented as $U_{\phi(x)}H^{\otimes n}$ with
\beq
U_{\phi(\vec{x})}=e^{i\sum_{Q\subset[n]}\phi_{Q}(\vec{x})\prod_{i\in Q}Z_{i}}, 
\eeq
where $\phi_{Q}(\vec{x}) \in \mathbb{R}$ are fixed functions to encode data in feature vector $\vec{x}$, and $Q$ a qubit subset. For instance, a typical example for two qubits and two features is $\phi_i(\vec{x}) = x_i$ and $\phi_{1,2}(\vec{x}) = (\pi - x_1)(\pi - x_2)$.  This example can be easily generalized to a larger number of features and qubits, with the resulting quantum circuit involving a large degree of quantum entanglement. 

Before moving forward with other methods, let us mention at this point that although quantum annealing approaches have also been proposed to implement quantum versions of SVM \cite{qsvmdwave}, the advantages each technology provides are very different. While the annealing approach promises an improvement in the generalization (which is a task in which SVMs already perform remarkably well), the proposed method with gate-based quantum computers offers better performance due to an easier classification task.

\subsection{QBoost on a quantum annealer}

First proposed by Neven et al. \cite{neven}, the QBoost algorithm is an ensemble model consisting on a set of weak Decision Tree classifiers, which are combined to form a strong classifier by finding the optimal binary weights. It does so via the minimization of the following cost function:
$$w^{opt}=arg min_{w}\Big(\sum_{s}^{S}\Big(\frac{1}{N}\sum_{i}^{N}w_{i}h_{i}(\vec{x}_{s})-y_{s}\Big)^{2}\Big)+\lambda\Vert w\Vert_{0},$$
where $h_{i}(\vec{x})\in[-1,1]$ is the $i$-th weak classifier, $S$ is the size of the dataset, $\vec{x}_{s}$ the datapoints and $y_{s}$ the labels which can be -1 or +1. The last term corresponds to a regularization term, and is controlled by the hyperparameter $\lambda$. Its goal is to favor weight sparsity, penalizing complex models with more weak classifiers in order to achieve a better generalization on unseen data.

\par Since they are irrelevant to the minimization problem, we can drop the constant terms, and the cost function can then be reformulated as
$$w^{opt} =argmin_{w}\Big(\frac{1}{N^{2}}\sum_{i,j}^{N}w_{i}w_{j}Corr(h_{i}, h_{j})$$
$$+\sum_{i}^{N}w_{i}(\lambda-\frac{2}{N}Corr(h_{i},y))\Big),$$
with $Corr(h_{i},h_{j})=\sum_{s}h_{i}(\vec{x}_{s})h_{j}(\vec{x}_{s})$ and $Corr(h_{i}, y)=\sum_{s}h_{i}(\vec{x}_{s})y_{s}$. In this formulation it is easy to see the following:
\begin{enumerate}
    \item Weak classifiers whose output is well correlated with the labels cause the bias term to be lowered through the term $Corr(h_{i},y)$.
    \item The quadratic part consists of the correlations between the weak classifiers, $Corr(h_{i}, h_{j})$.
    \item Classifiers that are strongly correlated with each other cause the energy to go up, thereby increasing the probability for one of them to be switched off.
    \item This agrees with the general paradigm of ensemble methods, which promotes a diversification of the classifiers in the ensemble for an improved generalization on new data.
\end{enumerate}
Weak classifiers are trained on the dataset sequentially, penalizing misclassified examples. We initialize a distribution $D_{1}(s)=\frac{1}{S}$, with $S$ the size of the dataset, and for every $h_{i}$ with $i=1\dots N$, we calculate:
$$\varepsilon_{i}=\sum_{s}D_{i}(s)\cdot h_{i}(x_{s})\hspace{5pt},\hspace{15pt}h_{i}(x_{s})\neq y_{s}$$
We then choose
$$w_{i}=\frac{1}{2}ln\Big(\frac{1-\varepsilon_{i}}{\varepsilon_{i}}\Big)$$
and update
$$D_{i+1}(s)=\frac{D_{i}(s)\cdot\exp(-w_{i}y_{s}h_{i}(x_{s}))}{Z_{i}}$$
where $Z_{i}$ is a normalization factor that ensures that $D_{i+1}$ is a distribution. The final strong classifier for a new point $x$ is then constructed as:
$$C(\vec{x})=sign\Big(\sum_{i}^{N}w_{i}h_{i}(\vec{x})-T\Big)$$
where $T$ is an optimal threshold that enhances results, computed as a post-processing step as in \cite{neven1}:
$$T=sign\Big(\frac{1}{S}\sum_{s}^{S}\frac{1}{N}\sum_{1}^{N}w_{i}^{opt}h_{i}(\vec{x})\Big)$$

This second formulation also gives a very useful representation of the problem at hand. Since the weights $w_{i}$ are binary variables, that is, $w_{i}^{2}=w_{i}$, one can write the function as
$$\sum_{i,j}Q_{i,j}w_{i}w_{j}$$
where $Q_{i,j}$ is a symmetrical (or equivalently, upper-diagonal) matrix that contains the coefficients $Corr(h_{i},h_{j})$ as its off-diagonal terms and $\frac{S}{N^{2}}+\lambda-2Corr(h_{i},y)$ in the diagonal. This formulation corresponds to a \textit{quadratic unconstrained binary optimization} (QUBO) problem, which is well-known to be NP-hard. As the number of weak classifiers grows, this function becomes exponentially harder to optimize classically, and hence opens the door to more efficient quantum methods.

\subsection{Classical algorithms}



Among Classical Deep Learning (DL) methods, a common way of detecting defects is based on the segmentation of these defects. Image segmentation consists in assigning a label to each pixel of the image so that pixels with the same label share certain characteristics, and has been applied in scenarios such as medical imaging, autonomous driving, etc. In this way, defect segmentation consists of separating the defects from the background, so that each type of defect will be assigned a different class and the remaining parts of the images will be associated to the same class. The task of detecting defects in images has been extensively studied during the last years in Computer Vision research and approaches have been proposed regarding to defect segmentation with DL models \cite{camilo} \cite{cagri}. Many of these approaches and reference architectures for defect detection have been based mainly in deep convolutional neural networks, due to their ability extracting image features \cite{ruoxu} \cite{donahue} \cite{tabernik} and they have shown impressive results. 
Although the classical DL algorithms we present here (SVM and Adaboost) are not the state of the art in this task, they are good references that have been used to do first approaches in defect detection \cite{ding}. It is important to remark at this point that the goal of this paper is to compare the performance of available quantum methods versus their classical counterparts, not to get the best possible performance in the task of detecting defects in images.
This way, we benchmarked the quantum methods presented in this paper against their classical counterparts, namely a linear and nonlinear SVM against the QSVM, and the AdaBoost algorithm against the QBoost. 
Support Vector Machines aim to find the hyperplane that best split the data into two regions, belonging each one to the dataset classes (in this case, normal data vs defect data). The margin is the distance between the nearest data point from each class and the hyperplane, and the goal is to find the hyperplane with the maximum margin. For linearly separable data, we can use a Linear SVM, which can find the optimal hyperplane built by a linear combination of all features that we have. In the case of a high dimensional, sparse problem with few irrelevant features, linear SVMs achieve great performance \cite{joachims}.  If the data is non-linearly separable, we have to transform the features to a new space using a function called kernel, that can transform non-linear to linear spaces where the data can be easily classified. In this work, we used a Radial Basis Function kernel (RBF) which uses the following function to transform the data:
$$\phi(x, center)=exp(-\gamma ||x -center||^2)$$
where gamma ($\gamma$) controls the influence of new features on the decision boundary. The higher the gamma, the more influence of the features will have on the decision boundary, and the curves generated will be more adapted to the existing data. If gamma is low, the curve will be broader and less fit to individual data points.

Adaptive Boosting, or AdaBoost, is a boosting technique that trains a sequence of weak models, each compensating the weaknesses of its predecessors. In this case we have used small decision trees, whose predictions are combined through a weighted majority vote to get the final prediction. At the beginning these weights are set to $w_{i}=1/N$, and for each successive iteration, the sample weights are individually modified and the learning algorithm is reapplied to the reweighted data. As iterations proceed, examples that are difficult to predict receive increasing influence. Each subsequent weak learner is thereby forced to concentrate on the examples that are missed by the previous ones in the sequence.
All these models were taken from Sci-kit Learn, and correspond to a LinearSVC, a non-linear NuSVC with RBF kernel ($nu=0.5$, $gamma=1/(n\_features*X.var()))$ \cite{scikitsvm}, and the AdaBoost \cite{scikitada} with a varying number of weak classifiers. The rest of the parameters were set as default.

\begin{table}
    \centering
    \begin{tabular}{|c|c|c|c|c|}
    \hline
        Model & Precision & Recall & F1 \\
        \hline
        \hline
        Linear SVM & 0.81 & 0.28 & 0.41 \\
        Non-Linear SVM & 0.87 & 0.57 & 0.69 \\
        AdaBoost(10 trees) & 0.71 & 0.68 & 0.70 \\
        AdaBoost (50 trees) & 0.76 & 0.84 & 0.80 \\
        \textbf{QSVM (simulated)} & \textbf{0.88} & \textbf{0.96} & \textbf{0.92} \\
        QBoost (9/10 trees, exhaustive) & 0.89 & 0.90 & 0.90 \\
        QBoost (9/10 trees, D-Wave) & 0.89 & 0.90 & 0.90 \\
        \hline
    \end{tabular}
    \caption{Results using PCA to reduce the dimension of the data to 10 features. Best classifier in boldface.}
    \label{tablepca10}
\end{table}

\section{Results}
\label{sec4}
The results presented here benchmark the algorithms for quantum and classical image classification, in three situations: (i) when including/not including dimensionality reduction via PCA, (ii) when enhancing the contrast of the images, and (iii) when varying the proportions in the dataset. Furthermore, we present a  detailed study of the fine-tuning of QBoost hyperparameters. Due to very large running times when using IBM quantum hardware mainly due to queuing issues (even with allocated time), we only present results obtained using Qiskit's statevector simulator.

\subsection{Dimensionality reduction}
Due to the low number of qubits available on current gate-based quantum hardware, reducing the dimension of the data is a must. Using PCA, we reduced the dimension of the data to 10 (Table \ref{tablepca10}) and 20 (Table \ref{tablepca20}) features (16 for the QSVM since the simulators can not go higher), and compared the performance of the different models.

\begin{table}
    \centering
    \begin{tabular}{|c|c|c|c|c|}
    \hline
        Model & Precision & Recall & F1 \\
        \hline
        \hline
        Linear SVM & 0.71 & 0.34 & 0.46 \\
        Non-Linear SVM & 0.90 & 0.56 & 0.69 \\
        AdaBoost (10 trees) & 0.85 & 0.75 & 0.80 \\
        AdaBoost (50 trees) & 0.76 & 0.73 & 0.74 \\
        \textbf{QSVM (16 qubits, simulated)} & \textbf{0.87} & \textbf{0.96} & \textbf{0.91} \\
        QBoost (7/10 trees, exhaustive) & 0.85 & 0.95 & 0.89 \\
        QBoost (7/10 trees, D-Wave) & 0.85 & 0.95 & 0.89 \\
        \hline
    \end{tabular}
    \caption{Results using PCA to reduce the dimension of the data to 20 features. Best classifier in boldface.}
    \label{tablepca20}
\end{table}

From the tables we can see how the quantum models already outperform their classical counterparts by a wide margin. It is worth mentioning also that quantum approaches have a slower running time due to latency problems when connecting to the QPUs and/or simulators online, as seen in Table \ref{tablepcaruntimes}. This issue, however, is not relevant since it does not scale up with the number of qubits, and is problem and hardware-independent.

If we do not apply PCA, we can no longer obtain results for the QSVM, but we can see in Table \ref{tableruntimes} how the running time of the QBoost is comparable to that of the classical AdaBoost when optimised using a D-Wave device. It is also notable how this time increases drastically when using a classical exhaustive optimiser for QBoost instead of D-Wave, highlighting the complexity of its cost function. It is also remarkable that the D-Wave solution is really close in precision to the one obtained for QBoost with a exhaustive optimization solver. These results are summarized in Table \ref{tablenopca}.

The improvement in performance for the QBoost seen in the above results becomes even more impressive when looking at the running times. Not only can it outperform in precision a classical AdaBoost with five times more trees, but also takes roughly the same amount of time as an AdaBoost with the same number of trees. The observed improvements in Qboost and QSVM with respect to their classical counterparts are a natural consequence of the fact that these algorithms use quantum effects, such as quantum paralelism and quantum entanglement, in order to explore pinpoint subtle patterns in the date in a more efficient way than classical methods.

\subsection{Contrast Enhancement}
\begin{figure}
    \centering
    \includegraphics[scale=0.36]{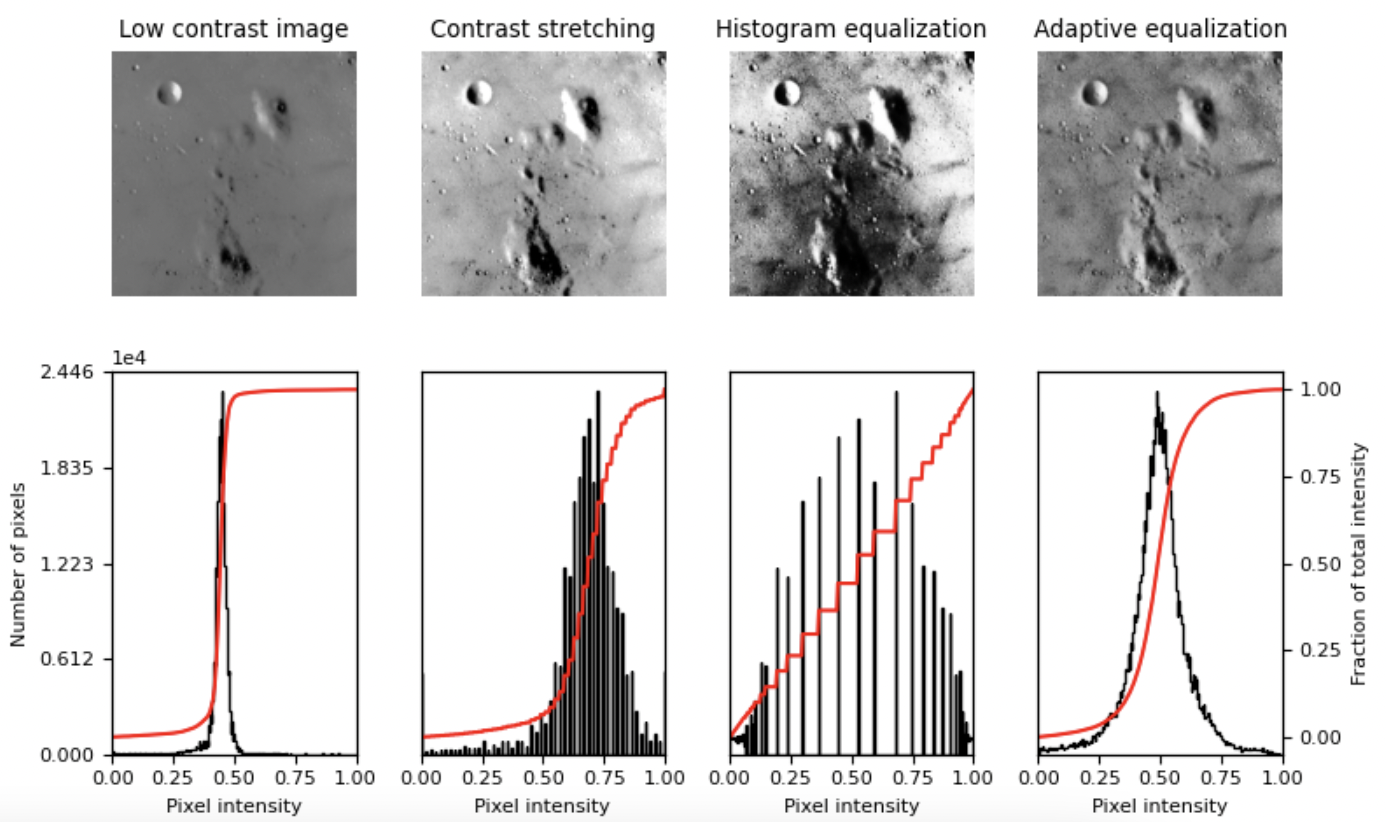}
    \caption{Example of the different contrast enhancement techniques used in this section, along with the transformation each of them produces on the image's histogram. Image from the Sci-kit image tutorial, see Ref.\cite{skimage}.}
    \label{contrast}
\end{figure}

Many of the images in the GDXray dataset contain low-contrast defects which are hard to locate even for the human eye (see Fig. \ref{defectfig}). This motivates the use of contrast enhancement techniques to try to highlight these defects. In this study, we have employed three techniques which are already implemented on the Sci-Kit Image framework, as shown in Fig. \ref{contrast}. After them, the same preprocessing that was explained in Section 1.2 was applied.

\begin{table}
    \centering
    \begin{tabular}{|c|c|c|}
    \hline
        Model & PCA 10 (sec) & PCA 20 (sec) \\
        \hline
        \hline
        \textbf{AdaBoost (10 trees)} & \textbf{0.093} & \textbf{0.138} \\
        AdaBoost (50 trees) & 0.301 & 0.569 \\
        QBoost (10 trees, exhaustive) & 1.543 & 1.725 \\
        QBoost (10 trees, D-Wave) & 12.588 & 20.104 \\
        \hline
    \end{tabular}
    \caption{Running times with dimensionality reduction to 10 and 20 features. Fastest classifier in boldface.}
    \label{tablepcaruntimes}
\end{table}
\begin{table}
    \centering
    \begin{tabular}{|c|c|}
    \hline
        Model & Runtime (min) \\
        \hline
        \hline
        AdaBoost (10 trees) & 6.2 \\
        AdaBoost (50 trees) & 49.0 \\
        QBoost (10 trees, exhaustive) & 113.0 \\
        \textbf{QBoost (10 trees, D-Wave)} & \textbf{6.0} \\
        \hline
    \end{tabular}
    \caption{Running times without dimensionality reduction. Fastest classifier in boldface.}
    \label{tableruntimes}
\end{table}
\begin{table}
    \centering
    \begin{tabular}{|c|c|c|c|c|}
    \hline
        Model & Precision & Recall & F1 \\
        \hline
        \hline
        Linear SVM & 0.73 & 0.77 & 0.75 \\
        Non-Linear SVM & 0.91 & 0.54 & 0.68 \\
        AdaBoost (10 trees) & 0.82 & 0.83 & 0.83 \\
        AdaBoost (50 trees) & 0.82 & 0.81 & 0.81 \\
        \textbf{QBoost (9/10 trees, exhaustive)} & \textbf{0.90} & \textbf{0.95} & \textbf{0.93} \\
        QBoost (9/10 trees, D-Wave) & 0.90 & 0.94 & 0.92 \\
        \hline
    \end{tabular}
    \caption{Results without using PCA. Best classifier in boldface.}
    \label{tablenopca}
\end{table}

\subsubsection{Contrast Stretching}
Contrast stretching (often called \emph{normalization}) is a simple image enhancement technique that attempts to improve the contrast in an image by ``stretching" the range of intensity values it contains to span a desired range of values, e.g., the the full range of pixel values that the image type concerned allows. It differs from the more sophisticated histogram equalization in that it can only apply a linear scaling function to the image pixel values. As a result the ``enhancement" is less harsh.

Before implementing the stretching, it is necessary to specify the upper and lower pixel value limits over which the image is to be normalized. Quite often these limits will just be the minimum and maximum pixel values that the image type concerned allows. For example for 8-bit greylevel images, the lower and upper limits might be 0 and 255, and in our case we chose the $2^{nd}$ and $98^{th}$ percentiles. Call the lower and the upper limits \emph{a} and \emph{b} respectively. The simplest sort of normalization then scans the image to find the lowest and highest pixel values currently present in the image. Call these \emph{c} and \emph{d}. Then each pixel \emph{P} is scaled using the following function:

$$P_{out}=(P_{in}-c)\Big(\frac{b-a}{d-c}\Big)+a$$

After applying this transformation to the images, the same simple preprocessing explained in the previous sections was used. The results for the different classifiers are detailed in Table \ref{nopcastretch}.
\begin{table}
    \centering
    \begin{tabular}{|c|c|c|c|c|}
    \hline
        Model & Precision & Recall & F1\\
        \hline
        \hline
        Linear SVM & 0.74 & 0.77 & 0.75 \\
        Non-Linear SVM & 0.90 & 0.54 & 0.67 \\
        AdaBoost(10 trees) & 0.77 & 0.72 & 0.75 \\
        AdaBoost (50 trees) & 0.75 & 0.83 & 0.79 \\
        \textbf{QBoost (9/10 trees, exhaustive)} & \textbf{0.9} & \textbf{0.92} & \textbf{0.91} \\
        QBoost (10/10 trees, D-Wave) & 0.92 & 0.86 & 0.89 \\
        \hline
    \end{tabular}
    \caption{Results after applying contrast stretching to the images. Best classifier in boldface.}
    \label{nopcastretch}
\end{table}

It is clear that this technique does not produce any significant changes in the performance of the models, when comparing to the results from the previous tables. Therefore we decided to use more complex techniques, which we explain in the following.

\subsubsection{Histogram Equalization}
Histogram equalization is an image processing technique used to improve contrast in images. By increasing the intensity range of the image, it effectively spreads out the range of intensities and therefore improves the visibility of low contrast features.

Let \emph{f} be an image represented as an \emph{mr} by \emph{mc} matrix of integer pixel intensities ranging from 0 to \emph{L}$-$1. \emph{L} is the number of possible intensity values, often 256. Let \emph{p} denote the normalized histogram of \emph{f} with a bin for each possible intensity. So

$$p_{n}=\frac{{\rm num.\hspace{3pt}pixels\hspace{3pt}with\hspace{3pt}intensity\hspace{3pt}} n}{{\rm total\hspace{3pt}num.\hspace{3pt}pixels}}\hspace{15pt}n=0,1,\dots L-1.$$

The histogram equalized image \emph{g} will be defined by

$$g_{i,j}=\left\lfloor(L-1)\sum_{n=0}^{f_{i,j}}p_{n}\right\rfloor,$$
where $\lfloor ~~ \rfloor$ rounds down to the nearest integer from below. This is equivalent to transforming the pixel intensities, \emph{k}, of \emph{f} by the function
$$T(k)=\left\lfloor(L-1)\sum_{n=0}^{k}p_{n}\right\rfloor.$$

Just like with contrast stretching, the preprocessing was also applied after this transformation. The results are summarized in Table \ref{nopcahisttable}.

\begin{table}
    \centering
    \begin{tabular}{|c|c|c|c|c|}
    \hline
        Model & Precision & Recall & F1 \\
        \hline
        \hline
        Linear SVM & 0.72 & 0.73 & 0.72 \\
        Non-Linear SVM & 0.91 & 0.53 & 0.67 \\
        AdaBoost(10 trees) & 0.78 & 0.72 & 0.75 \\
        AdaBoost (50 trees) & 0.81 & 0.79 & 0.80 \\
        QBoost (10/10 trees, exhaustive) & 0.91 & 0.88 & 0.89 \\
        \textbf{QBoost (9/10 trees, D-Wave)} & \textbf{0.86} & \textbf{0.95} & \textbf{0.90} \\
        \hline
    \end{tabular}
    \caption{Results after applying histogram equalization to the images. Best classifier in boldface.}
    \label{nopcahisttable}
\end{table}

Overall we noticed that the running time increases for almost every model due to the more complex non-linearity of the transformation. Notice also that, in this case, the D-Wave solution was even better than the one obtained by the exhaustive solver for QBoost, showing that the minimum of the QBoost cost function does not necessarily correspond to the maximums of the figures of merit considered in the table. In any case, the obtained results are again quite similar to those without any enhancement technique. Hence, we decided to discard this method and try a third one.

\subsubsection{Adaptative Equalization}
Similar to histogram equalization, this technique applies the exact same transformation, but instead of transforming the image as a whole, it does so by modifying different histograms. These histograms correspond to different areas of the image, effectively applying this enhancement in areas where it is most needed. Since our images mostly show low gradient contrast around the defects, but high contrast everywhere else, it is easy to see why these technique might be the most appropriate for our case. Again, the preprocessing procedure was applied after this transformation. The results are shown in Table \ref{nopcaadaptable}. We have observed that the increase in running time is even more significant than that for histogram equalization. 

\begin{table}
    \centering
    \begin{tabular}{|c|c|c|c|c|}
    \hline
        Model & Precision & Recall & F1 \\
        \hline
        \hline
        Linear SVM & 0.76 & 0.74 & 0.75 \\
        Non-Linear SVM & 0.92 & 0.56 & 0.70 \\
        AdaBoost(10 trees) & 0.73 & 0.77 & 0.75 \\
        AdaBoost (50 trees) & 0.77 & 0.75 & 0.76 \\
        QBoost (9/10 trees, exhaustive) & 0.85 & 0.94 & 0.89 \\
        \textbf{QBoost (9/10 trees, D-Wave)} & \textbf{0.86} & \textbf{0.96} & \textbf{0.91} \\
        \hline
    \end{tabular}
    \caption{Results after applying adaptative equalization to the images. Best classifier in boldface.}
    \label{nopcaadaptable}
\end{table}

Overall, what we see is that contrast enhancement techniques do not turn out to result in a valuable improvement of the results. Quantum classifiers are already performing better than their classical counterparts without applying these techniques, and do not show a significant increase in the precision measures after adjusting contrast whereas the runtime increases significantly. Because of this, we have opted not to continue using them for the rest of this work, though we believe that they should be considered for other datasets in the context of a wider framework for quantum artificial vision. 

\subsection{QBoost fine-tuning}
As we have seen in the previous sections, the QBoost algorithm has proven to be the best performing of the models, showing significantly better performance than its classical counterpart with and without dimensionality reduction. This remarking improvement has led us to focus on the exploration of this model, and during this section we present results of both performance and inference time for this model while varying its different parameters.

\subsubsection{Regularisation}
QBoost is an ensemble model consisting on several Decision Tree Classifiers (DTC). The weight of each DTC towards the final strong classifier’s output is calculated via the optimisation of the QUBO cost function. The regularisation parameter, $\lambda$, only affects this optimisation process, which means we can train the DTCs and then look for the optimal $\lambda$ using D-Wave’s Hybrid Sampler without having to wait for the DTCs to be trained on every iteration. This means the running time for this exploration is significantly lower than that for its classical counterpart.

Another aspect worth considering of this dataset is the unbalance between the classes, which prompts for the use of balancing techniques such as Random Under Sampling (RUS). The results in Table \ref{norus10table} and Table \ref{rus10table} show the performance of the QBoost model when varying its regularisation parameter with and without RUS (50-50 sampling strategy) respectively. Since the metrics do not change unless the final number of classifiers changes, we only show results for the lowest regularisation value that leads to each number of final DTCs.

\begin{table}
    \centering
    \begin{tabular}{|c|c|c|c|c|}
    \hline
        Regularisation & Number of classifiers & Precision & Recall & F1\\
        \hline
        \hline
        0 & 10/10 & 0.95 & 0.84 & 0.90\\
        \textbf{5} & \textbf{9/10} & \textbf{0.90} & \textbf{0.93} & \textbf{0.92}\\
        25 & 8/10 & 0.90 & 0.90 & 0.90\\
        45 & 7/10 & 0.87 & 0.95 & 0.91\\
        60 & 6/10 & 0.89 & 0.91 & 0.90\\
        75 & 5/10 & 0.88 & 0.93 & 0.91\\
        85 & 4/10 & 0.89 & 0.89 & 0.89\\
        100 & 4/10 & 0.89 & 0.89 & 0.89\\
        \hline
    \end{tabular}
    \caption{Performance results with different regularisation parameters. The running time for each iteration is around 20 seconds. Best result in boldface.}
    \label{norus10table}
\end{table}
\begin{table}
    \centering
    \begin{tabular}{|c|c|c|c|c|}
    \hline
        Regularisation & Number of classifiers & Precision & Recall & F1\\
        \hline
        \hline
        0 & 10/10 & 0.97 & 0.68 & 0.80\\
        20 & 9/10 & 0.94 & 0.80 & 0.87\\
        35 & 8/10 & 0.95 & 0.77 & 0.85\\
        45 & 6/10 & 0.94 & 0.78 & 0.85\\
        \textbf{65} & \textbf{5/10} & \textbf{0.91} & \textbf{0.87} & \textbf{0.89}\\
        75 & 3/10 & 0.86 & 0.90 & 0.88\\
        85 & 2/10 & 0.94 & 0.63 & 0.75\\
        100 & 2/10 & 0.94 & 0.63 & 0.75\\
        \hline
    \end{tabular}
    \caption{Performance results with different regularisation parameters after applying RUS to the dataset. The running time for each iteration is around 20 seconds. Best result in boldface.}
    \label{rus10table}
\end{table}

\begin{figure}
    \centering
    \includegraphics[scale=0.21]{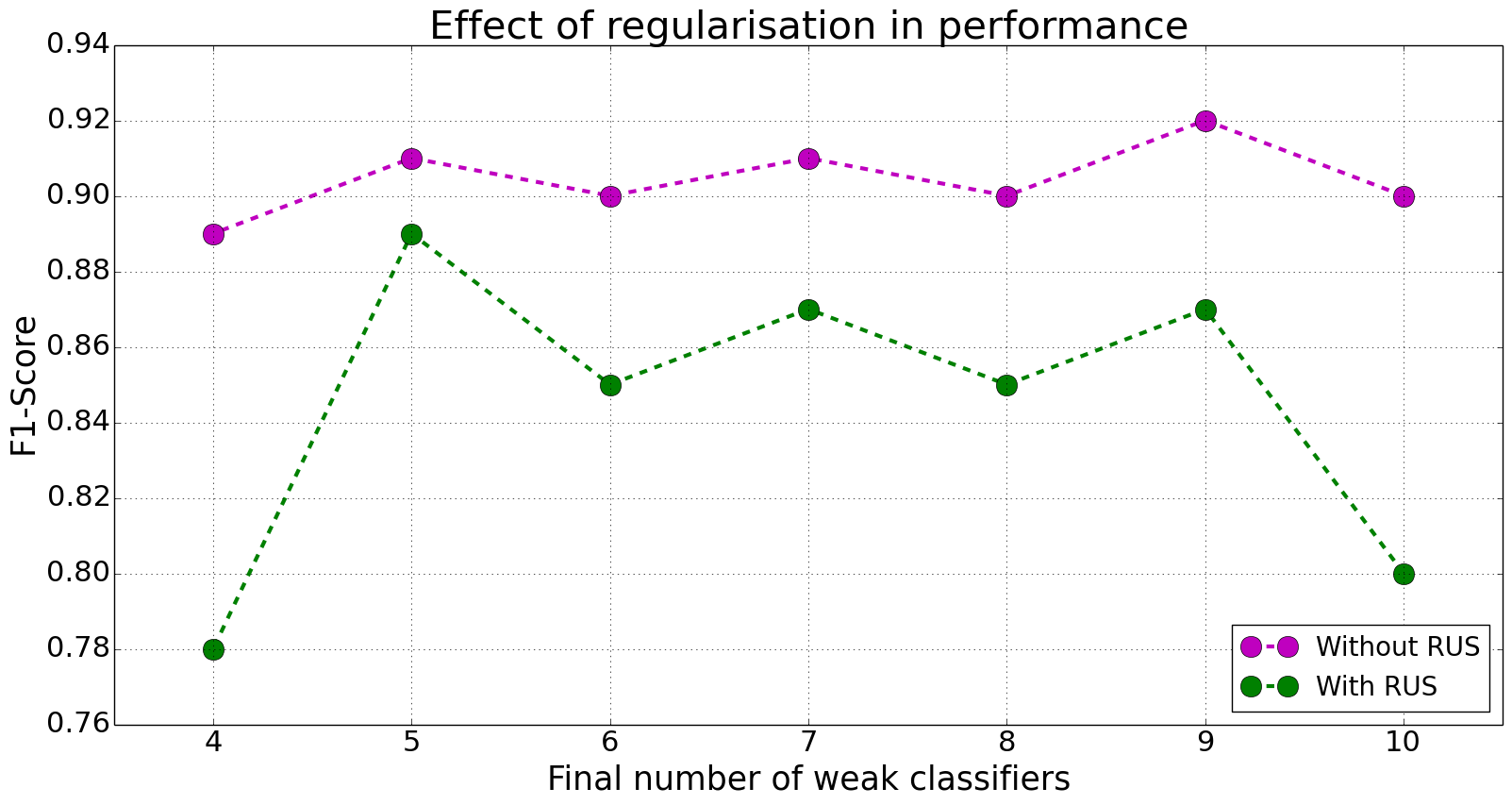}
    \caption{[Color online] F1 score results for different regularisation parameter values with and without RUS, with an initial number of 10 weak classifiers.}
    \label{regularisation10}
\end{figure}

When applying RUS, one can see how the increase in regularisation leads to a quicker decrease in the final number of DTCs. This means that balancing the dataset induces correlations between the DTCs, which in turn favours weight sparsity in the cost function, causing more classifiers to be switched off. We can also observe how the model tends to overfit when using this technique, and the difference in performance is shown in Fig. \ref{regularisation10}.

\begin{figure}
    \centering
    \includegraphics[scale=0.21]{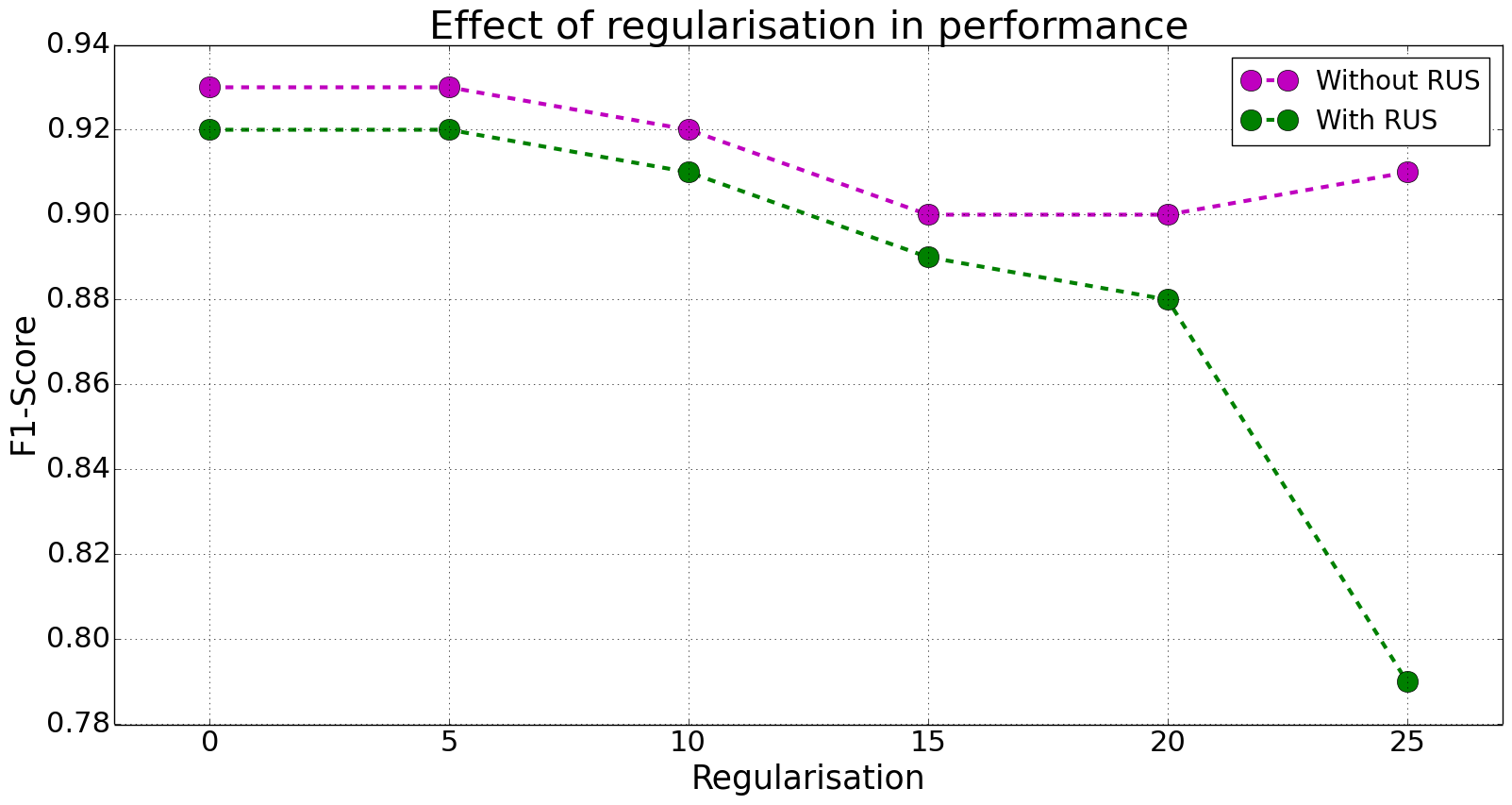}
    \caption{[Color online] F1 score results for different regularisation parameter values with and without RUS, with an initial number of 50 weak classifiers.}
    \label{regularisation50}
\end{figure}

Although adding RUS seems to induce some instability in the performance of the model, the remarkable speed at which the model is trained allows us to use a higher initial number of DTCs. This results are summarized in Table \ref{norus50table} and Table \ref{rus50table}, and in Figure \ref{regularisation50} we can see how increasing the number of DTCs maintains the stability in performance when using RUS.

\begin{table}
    \centering
    \begin{tabular}{|c|c|c|c|c|}
    \hline
        Regularisation & Number of classifiers & Precision & Recall & F1\\
        \hline
        \hline
        0 & 47/50 & 0.90 & 0.96 & 0.93\\
        \textbf{5} & \textbf{43/50} & \textbf{0.89} & \textbf{0.98} & \textbf{0.93}\\
        10 & 38/50 & 0.88 & 0.97 & 0.92\\
        15 & 31/5 & 0.83 & 0.98 & 0.90\\
        20 & 21/50 & 0.83 & 0.98 & 0.90\\
        25 & 14/50 & 0.84 & 0.98 & 0.91\\
        30 & 9/50 & 0.80 & 0.98 & 0.89\\
        35 & 6/50 & 0.84 & 0.95 & 0.89\\
        40 & 3/50 & 0.82 & 0.97 & 0.89\\
        45 & 2/50 & 0.87 & 0.85 & 0.86\\
        50 & 1/50 & 0.83 & 0.92 & 0.87\\
        \hline
    \end{tabular}
    \caption{Performance results with different regularisation parameters. The running time for each iteration is around 45 seconds. Best result in boldface.}
    \label{norus50table}
\end{table}
\begin{table}
    \centering
    \begin{tabular}{|c|c|c|c|c|}
    \hline
        Regularisation & Number of classifiers & Precision & Recall & F1\\
        \hline
        \hline
        \textbf{0} & \textbf{50/50} & \textbf{0.96} & \textbf{0.89} & \textbf{0.92}\\
        5 & 48/50 & 0.95 & 0.89 & 0.92\\
        10 & 35/50 & 0.94 & 0.89 & 0.91\\
        15 & 14/50 & 0.90 & 0.87 & 0.89\\
        20 & 5/50 & 0.85 & 0.91 & 0.88\\
        25 & 1/50 & 0.87 & 0.73 & 0.79\\
        \hline
    \end{tabular}
    \caption{Performance results with different regularisation parameters after applying RUS to the dataset. The running time for each iteration is around 30 seconds. Best result in boldface.}
    \label{rus50table}
\end{table}

We can see how, even though applying RUS requires a more careful exploration of the regularisation parameter due to the faster decrease in the number of DTCs, the performance is equally stable and overall very similar to that without it. Since there is no observable advantage, we have decided to continue without using this technique either.

\subsubsection{Inference Time}
Current deep learning approaches to computer vision require really complex models to implement the required task \cite{camilo, cagri, ruoxu, donahue, tabernik, ding, joachims}, in turn causing a long time to classify an image. This is highly detrimental when using these models in a real scenario (e.g., real-time video image classification), and makes it difficult to implement them in a production environment. In this section we have explored in depth how each parameter of our model affects this inference time, and how to find a good trade-off between performance and speed.

Using the 546 images of our test set, we increased the regularisation parameter for different DTC depths. The inference times shown in Tables \ref{infdepth2}, \ref{infdepth3} and \ref{infdepth4} were averaged over the regularisation values that give the same final number of weak classifiers.
\bigskip
\begin{table}
    \centering
    \begin{tabular}{|c|c|c|c|c|c|}
    \hline
        Regul. & Num. classif. & Precision & Recall & F1 & Avg. inf. time\\
        \hline
        \hline
        \textbf{0} & \textbf{9/10} & \textbf{0.86} & \textbf{0.96} & \textbf{0.91} & \textbf{1.41}\\
        44 & 8/10 & 0.87 & 0.94 & 0.90 & 1.82\\
        50 & 7/10 & 0.83 & 0.96 & 0.89 & 1.83\\
        56 & 5/10 & 0.81 & 0.97 & 0.88 & 1.86\\
        72 & 4/10 & 0.83 & 0.95 & 0.89 & 1.31\\
        80 & 3/10 & 0.77 & 0.98 & 0.86 & 1.63\\
        88 & 2/10 & 0.82 & 0.85 & 0.84 & 1.40\\
        \hline
    \end{tabular}
    \caption{Performance results with different regularisation parameters and a tree depth of 2. The average inference time is in seconds. Best result in boldface.}
    \label{infdepth2}
\end{table}

\begin{table}
    \centering
    \begin{tabular}{|c|c|c|c|c|c|}
    \hline
        Regul. & Num. classif. & Precision & Recall & F1 & Avg. inf. time \\
        \hline
        \hline
        0 & 10/10 & 0.95 & 0.84 & 0.90 & 1.592\\
        \textbf{4} & \textbf{9/10} & \textbf{0.90} & \textbf{0.93} & \textbf{0.92} & \textbf{1.392}\\
        26 & 8/10 & 0.90 & 0.90 & 0.90 & 1.237\\
        44 & 7/10 & 0.87 & 0.95 & 0.91 & 1.450\\
        60 & 6/10 & 0.89 & 0.91 & 0.90 & 1.226\\
        74 & 5/10 & 0.88 & 0.93 & 0.91 & 1.308\\
        84 & 4/10 & 0.89 & 0.89 & 0.89 & 1.649\\
        \hline
    \end{tabular}
    \caption{Performance results with different regularisation parameters and a tree depth of 3. The average inference time is in seconds. Best result in boldface.}
    \label{infdepth3}
\end{table}

\begin{table}
    \centering
    \begin{tabular}{|c|c|c|c|c|c|}
    \hline
        Regul. & Num. classif. & Precision & Recall & F1 & Avg. inf.   time \\
        \hline
        \hline
        0 & 10/10 & 0.94 & 0.89 & 0.91 & 2.696\\
        6 & 9/10 & 0.91 & 0.94 & 0.92 & 1.856\\
        54 & 8/10 & 0.92 & 0.91 & 0.91 & 1.718\\
        \textbf{74} & \textbf{7/10} & \textbf{0.89} & \textbf{0.97} & \textbf{0.93} & \textbf{1.310}\\
        88 & 6/10 & 0.92 & 0.92 & 0.92 & 1.781\\
        118 & 5/10 & 0.88 & 0.97 & 0.92 & 1.393\\
        124 & 4/10 & 0.90 & 0.89 & 0.90 & 1.639\\
        142 & 3/10 & 0.86 & 0.95 & 0.91 & 2.032\\
        \hline
    \end{tabular}
    \caption{Performance results with different regularisation parameters and a tree depth of 4. The average inference time is in seconds. Best result in boldface.}
    \label{infdepth4}
\end{table}
\begin{figure}
    \centering
    \includegraphics[scale=0.21]{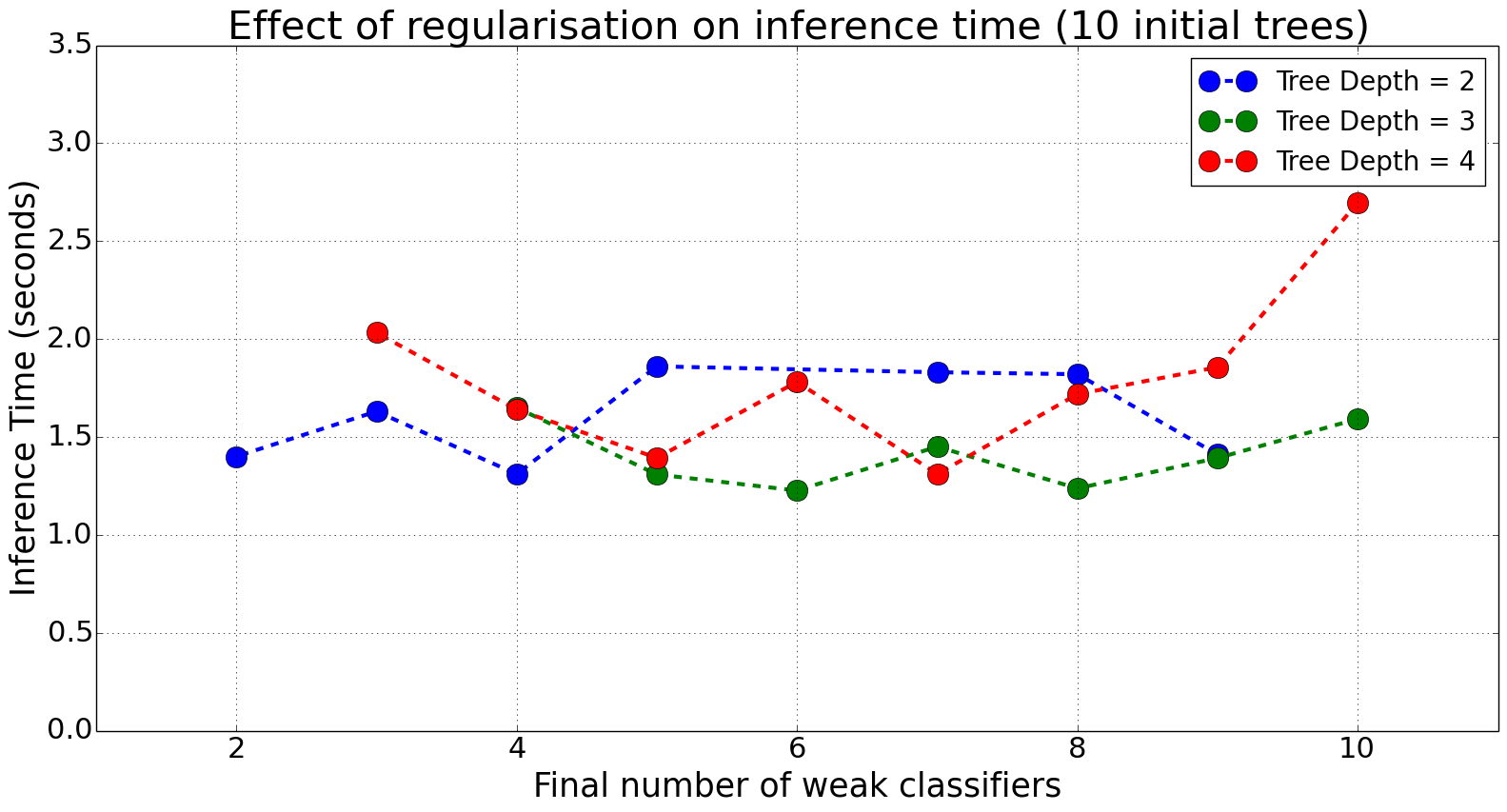}
    \caption{[Color online] Inference time for different regularisation parameter values, with an initial number of 10 weak classifiers and different tree depths.}
    \label{inference10}
\end{figure}

\begin{figure}
    \centering
    \includegraphics[scale=0.21]{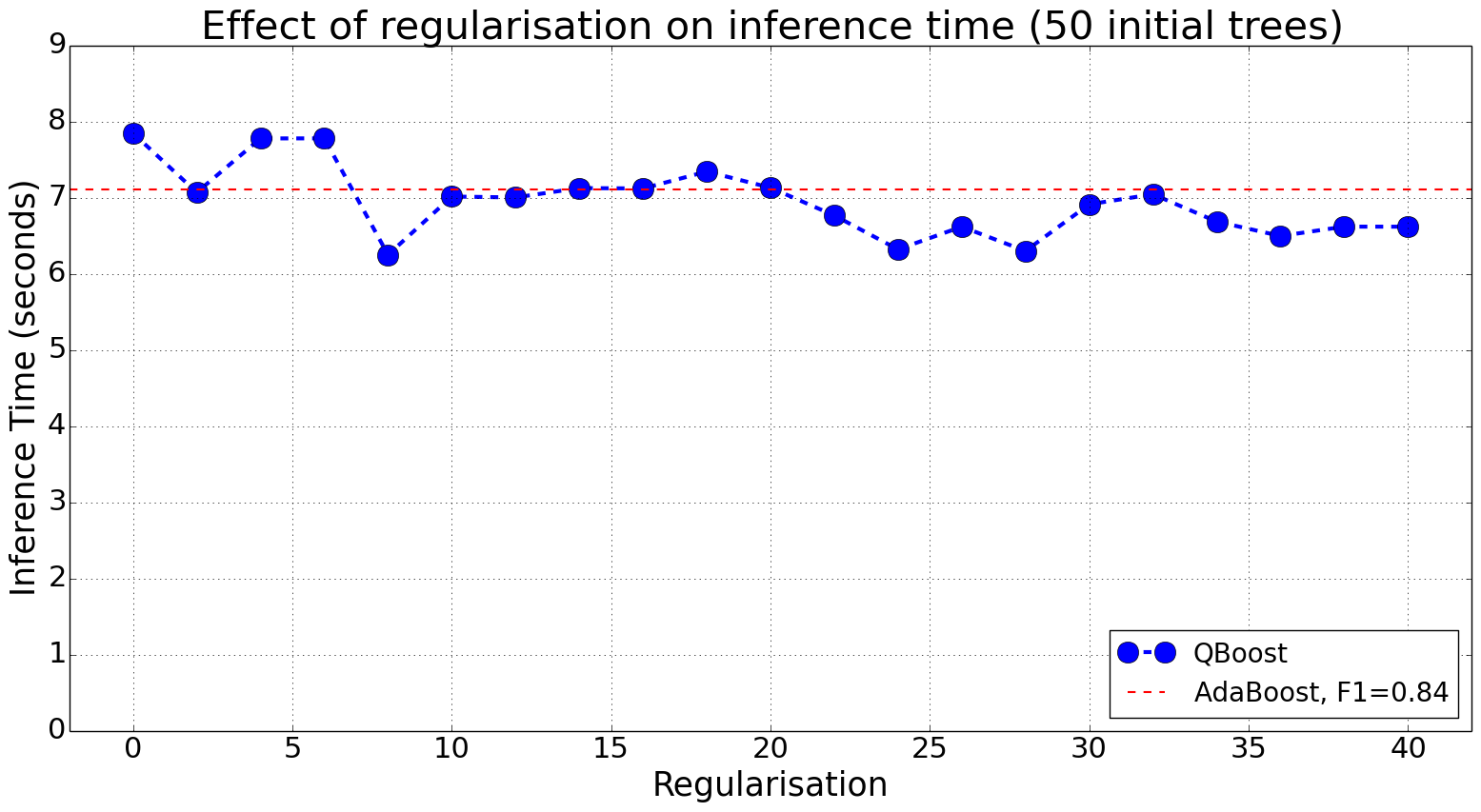}
    \caption{[Color online] Inference time for different regularisation parameter values, with an initial number of 50 weak classifiers and a tree depth of 4.}
    \label{inference50}
\end{figure}

\begin{table}
    \centering
    \begin{tabular}{|c|c|c|c|c|c|}
    \hline
        Regul. & Num. classif. & Precision & Recall & F1 & Inf. time\\
        \hline
        \hline
        0 & 48/50 & 0.94 & 0.94 & 0.94 & 7.845\\
        2 & 47/50 & 0.94 & 0.95 & 0.94 & 7.067\\
        4 & 45/50 & 0.93 & 0.95 & 0.94 & 7.776\\
        6 & 45/50 & 0.93 & 0.95 & 0.94 & 7.777\\
        8 & 44/50 & 0.93 & 0.95 & 0.94 & 6.248\\
        10 & 42/50 & 0.93 & 0.95 & 0.94 & 7.013\\
        \textbf{12} & \textbf{39/50} & \textbf{0.91} & \textbf{0.97} & \textbf{0.94} & \textbf{7.001}\\
        14 & 37/50 & 0.90 & 0.97 & 0.93 & 7.124\\
        16 & 34/50 & 0.90 & 0.96 & 0.93 & 7.119\\
        18 & 31/50 & 0.88 & 0.98 & 0.93 & 7.341\\
        20 & 27/50 & 0.88 & 0.98 & 0.93 & 7.129\\
        22 & 22/50 & 0.89 & 0.98 & 0.93 & 6.765\\
        24 & 21/50 & 0.87 & 0.98 & 0.92 & 6.319\\
        26 & 19/50 & 0.87 & 0.98 & 0.92 & 6.617\\
        28 & 16/50 & 0.88 & 0.97 & 0.93 & 6.298\\
        30 & 14/50 & 0.89 & 0.97 & 0.93 & 6.911\\
        32 & 13/50 & 0.87 & 0.98 & 0.93 & 7.043\\
        34 & 11/50 & 0.87 & 0.97 & 0.92 & 6.684\\
        36 & 9/50 & 0.87 & 0.97 & 0.92 & 6.495\\
        38 & 8/50 & 0.88 & 0.97 & 0.92 & 6.618\\
        40 & 7/50 & 0.86 & 0.98 & 0.91 & 6.617\\
        \hline
    \end{tabular}
    \caption{Performance results with different regularisation parameters, a tree depth of 4 and 50 initial weak classifiers. The  inference time is in seconds. Best result in boldface.}
    \label{inference50table}
\end{table}

We can see in Fig. \ref{inference10} that the inference time remains relatively constant with respect to the regularisation parameters and tree depth, with an average time of 2.94 miliseconds per image. Since these two hyperparameters do not seem to affect the inference time, the last step is trying to increase the initial number of classifiers. Table \ref{inference50table} shows the results for the QBoost classifier with a tree depth of 4 and 50 initial classifiers. It can be better visualized in Fig. \ref{inference50}, including a comparison with the classical AdaBoost (and its performance).

\begin{figure}
    \centering
    \includegraphics[scale=0.21]{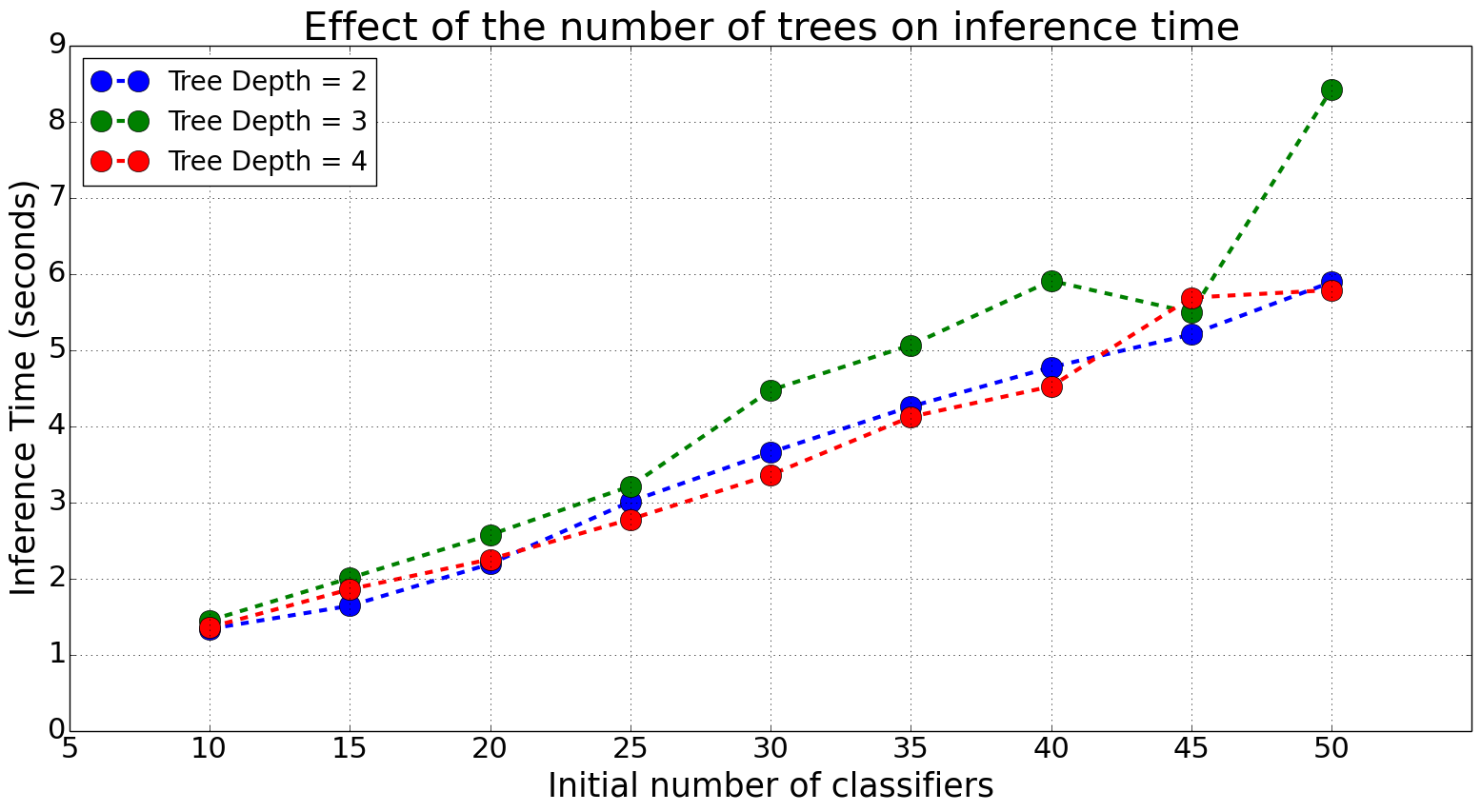}
    \caption{[Color online] Inference time for different initial number of DTCs, with a fixed regularisation value of 19 and different tree depths.}
    \label{inferencelinear}
\end{figure}

From the results it is clear that the initial number of DTCs affects the inference time, which seems to evolve similarly to that of the AdaBoost classifier but with a much better performance. To better understand this dependence, Fig. \ref{inferencelinear} seems to indicate the existence of a linear evolution with respect to the initial number of weak classifiers, which in turn eases the study of the trade-off between speed and performance.

\section{Conclusions and outlook}
\label{sec5}
In this paper we have benchmarked two quantum algorithms for artificial vision on NISQ devices: a quantum SVM on universal gate-based quantum computers, and a QBoost algorithm on D-Wave's quantum annealer. We have seen that these quantum algorithms outperform in precision their classical counterparts. Even with drastic dimensionality reduction, both the QSVM and the QBoost algorithms give impressive results, and we expect the performance of the former to improve in the near future with the evolution of gate-based quantum hardware. In the case of the QBoost, it has shown that current quantum machine learning models can already outperform classical solvers while maintaining training and inference times, which highlights the current capabilities of quantum annealing technologies. What is more, while QSVM requires a quantum computer for performing both training and inference tasks, QBoost-based solution is able to perform model training in a quantum system and then deploy the model in a classic system. This makes it possible the exploitation of quantum-based methods in real industrial  manufacturing scenarios in the near future, without requiring to adapt the ecosystem to use deploy and use them. And this is even more relevant, given the low energy consumption of quantum processors compared to that of traditional HPC infrastructures. As future work, we see the possibility of implementing benchmarks against CNN-based computer vision algorithms, both for qantum and quantum-inspired approaches. Our results show that quantum machine learning has a bright future when applied to real-life problems, and specifically to computer vision tasks. 

\bigskip 
{\bf Acknowledgements.-} DG, VO and RO acknowledge discussions with the rest of the technical team at Multiverse Computing as well as Creative Destruction Lab, BIC-Gipuzkoa, DIPC, Ikerbasque, Diputaci\'on de Gipuzkoa and Basque Government for constant support. We would like to acknowledge the constructive discussion with the technical team at Data Analytics and Artificial Intelligence research group at IKERLAN. Special thanks to Mikel Azkarate-Askatsua, Jokin Labaien, Samuel Vidal and Josu Ircio. This work was supported in part by the Basque Government through the BRTA Quantum project under the ELKARTEK program.

\end{document}